\documentclass[aps, pra, twocolumn, superscriptaddress, nofootinbib, amsmath,amssymb]{revtex4-2}

\usepackage{graphicx}
\usepackage{braket} 
\usepackage{physics} 

\usepackage{xcolor}

\begin{document}

\title{Benchmarking Single-Qubit Gates on a Neutral Atom\\ Quantum Processor}

\author{A.~Rozanov}
\email{a.rozanov61@gmail.com}
\affiliation{Quantum Technology Centre and Faculty of Physics, M.V. Lomonosov Moscow State University, 1 Leninskie Gory, Moscow, 119991, Russia}

\author{B.~Bantysh}
\affiliation{Russian Quantum Center, 30 Bolshoy Boulevard, building 1, Moscow, 121205, Russia}
\affiliation{NRC ``Kurchatov Institute'', Moscow 123182, Russia}

\author{I.~Bobrov}
\affiliation{Quantum Technology Centre and Faculty of Physics, M.V. Lomonosov Moscow State University, 1 Leninskie Gory, Moscow, 119991, Russia}

\author{G.~Struchalin}
\affiliation{Quantum Technology Centre and Faculty of Physics, M.V. Lomonosov Moscow State University, 1 Leninskie Gory, Moscow, 119991, Russia}

\author{S.~Straupe}
\affiliation{Quantum Technology Centre and Faculty of Physics, M.V. Lomonosov Moscow State University, 1 Leninskie Gory, Moscow, 119991, Russia}
\affiliation{Russian Quantum Center, 30 Bolshoy Boulevard, building 1, Moscow, 121205, Russia}

\date{\today}

\begin{abstract}
We present benchmarking results for single-qubit gates implemented on a neutral atom quantum processor using Direct Randomized Benchmarking (DRB) and Gate Set Tomography (GST). The DRB protocol involves preparing stabilizer states, applying $m$ layers of native single-qubit gates, and measuring in the computational basis, providing an efficient error characterization under a stochastic Pauli noise model. GST enables the full, self-consistent reconstruction of quantum processes, including gates, input states, and measurements. Both protocols provide robust to state preparation and measurement (SPAM) errors estimations of gate performance, offering complementary perspectives on quantum gate fidelity. For single-qubit gates, DRB yields an average fidelity of $99.963 \%$. The protocol was further applied to a 25-qubit array under global single-qubit control. GST results are consistent with those obtained via DRB. We also introduce a gauge optimization procedure for GST that brings the reconstructed gates, input states, and measurements into a canonical frame, enabling meaningful fidelity comparisons while preserving physical constraints. These constraints of the operators -- such as complete positivity and trace preservation -- are enforced by performing the optimization over the Stiefel manifold. The combined analysis supports the use of complementary benchmarking techniques for characterizing scalable quantum architectures.
\end{abstract}

\maketitle

\section{Introduction}

Quantum computers represent a promising platform for high-performance computational tasks~\cite{ref:QQ_REV_1, ref:QQ_REV_2, ref:QQ_REV_3}. Multiple physical implementations of qubits have been developed, including systems based on cold neutral atoms~\cite{ref:phys_realiz_1,ref:phys_realiz_2}, trapped ions~\cite{ref:phys_realiz_3,ref:phys_realiz_4}, photonic devices~\cite{ref:phys_realiz_5, ref:phys_realiz_6}, and superconducting circuits~\cite{ref:phys_realiz_7, ref:phys_realiz_8}. Despite significant progress, practical deployment is hindered by decoherence and operational errors~\cite{ref:quantum_noise_1, ref:quantum_noise_2}, which necessitate comprehensive diagnostic techniques and hardware optimization strategies~\cite{ref:bench_rev_1}. Accurate preparation, manipulation, and measurement of quantum states are essential components of quantum computation and must be precisely characterized to ensure reliable operation.

In this work, we focus on a platform based on cold neutral atoms. Specifically we use qubits encoded in hyperfine sublevels of the electronic ground state of rubidium atoms, which are cooled in a magneto-optical trap and loaded into an array of optical tweezers. Achieving fast and high-fidelity gate operations is essential for the advancement of this architecture.

Beyond conventional methods -- such as quantum state, process, and detector tomography~\cite{ref:q_tomo_1, ref:q_tomo_2, ref:q_tomo_3, ref:q_tomo_4}, alternative protocols have emerged to overcome limitations in scalability and the assumption of perfectly accurate measurements, which is incompatible with the state-preparation and measurement (SPAM) errors inherent to physical devices~\cite{ref:inf_compl_1, ref:inf_compl_2, ref:RB_1, ref:GST_1, ref:GST_2}. Notable examples include randomized benchmarking (RB)~\cite{ref:RB_1, ref:RB_2, ref:RB_3, ref:RB_4}, spectral channel benchmarking~\cite{ref:CSB_1}, and gate set tomography (GST)~\cite{ref:GST_1, ref:GST_2, ref:GST_3}.

In this paper, we apply Direct RB (DRB) and GST protocols to benchmark single-qubit gates on a neutral atom quantum processor. We demonstrate an average DRB fidelity of $99.963 ^{+0.015}_{-0.013}\%$ for single-qubit operations and extend the protocol to a 25-qubit array under global control, observing no significant degradation. Additionally, we introduce a gauge optimization method for GST based on joint diagonalization and fidelity maximization. Physical constraints of the operators, such as complete positivity and trace preservation, are maintained by performing the optimization over the Stiefel manifold. We also develop a novel two-parameter calibration scheme embedded directly into the DRB protocol, enabling extraction of coherent control errors without auxiliary measurements. Our results validate the accuracy and diagnostic utility of these benchmarking techniques for neutral atom architectures.

The structure of the paper is as follows. Section~\ref{sec:background} outlines the theoretical background and formalisms. Section~\ref{sec:simulation} discusses numerical simulations using a well-established error model based on dephasing, and introduces the GST calibration approach. Section~\ref{sec:experiment} presents experimental results from a neutral atom quantum processor, including the DRB-based calibration that improves gate fidelity.

\section{\label{sec:background}Background}

\subsection{Mathematical Formalism}

Our theoretical framework for benchmarking relies on the superoperator formalism. Let \( U \) be a unitary operator acting on an initial quantum state with density matrix \( \rho_{\text{in}} \). The corresponding superoperator \( G_U \) is defined as~\cite{ref:NCQ}:
\begin{equation}
    G_U = U \otimes U^*,
\end{equation}
where \( U^* \) denotes the complex conjugate of \( U \). This construction allows the density matrix to be reshaped into a \( d^2 \)-dimensional column vector \( \vec{\rho} \), with \( d = 2^n \) for a system of \( n \) qubits. The action of the channel then becomes:
\begin{equation}
    \vec{\rho}_{\text{out}} = G_U \vec{\rho}_{\text{in}}.
\end{equation}

By vectorizing the density matrix, quantum evolution is recast as a linear map acting on a vector space, allowing for a more structured and readily analyzable formulation. However, actual physical processes deviate from unitary evolution due to interactions with the environment. Typical types of noise include bit-flip and phase-flip errors, depolarization, and amplitude damping~\cite{ref:NCQ, ref:ErMods}. For neutral atom platforms, an additional significant error source is atom loss from the trap.

A noisy quantum process can be represented by a superoperator \( G_E \) constructed from the Kraus operators \( \{ E_k \} \) of the corresponding quantum channel:
\begin{equation}
    G_E = \sum_k E_k \otimes E_k^*. \label{eq:kraus_superoperator}
\end{equation}

Because the exact noise model is generally not known in experiments, gate performance is evaluated using fidelity metrics. Two common measures are the entanglement fidelity \( F^{ent} \) and the average gate fidelity $F^{\text{avg}}$~\cite{ref:RB_3, ref:fid_form_1}:
\begin{equation}
    F^{ent} = \frac{\mathrm{Tr}(G_U^\dagger \tilde{G})}{d^2}, 
\label{eq:ent_fid}
\end{equation}
\begin{equation}
    F^{\text{avg}} = \frac{d F^{ent} + 1}{d + 1}.
\label{eq:fid_a_e_couple}
\end{equation}

Here, \( \tilde{G} \) denotes the experimentally realized (noisy) superoperator, and \( G_U \) is the ideal one. In numerical simulations where both are known explicitly, these metrics enable direct comparison of theoretical and empirical performance, serving as a consistency check for the adopted noise model.

\subsection{Randomized benchmarking}

Randomized benchmarking (RB) provides a scalable method for quantifying the fidelity of quantum gate operations while mitigating sensitivity to SPAM (state preparation and measurement) errors~\cite{ref:RB_1, ref:RB_2}. The method measures the decay of the success probability as a function of the circuit depth. In this work, we adopt the direct randomized benchmarking (DRB) protocol~\cite{ref:RB_5, ref:RB_6}.

We use the following notation throughout this section: $m$ denotes the number of gate layers in a circuit, and 
$K$ is the number of randomized circuits per depth. The superoperators $G_i$ correspond to gates sampled from a generating set of the Clifford group. The operators \( G_I \) and \( G_M \) represent the initialization and measurement operations, which prepare a stabilizer state and map the output to the computational basis, respectively. Their noisy implementations are denoted by \( \tilde{G} \).

\begin{figure}[t!]
	\centering  
	\includegraphics[width = 0.51\textwidth]{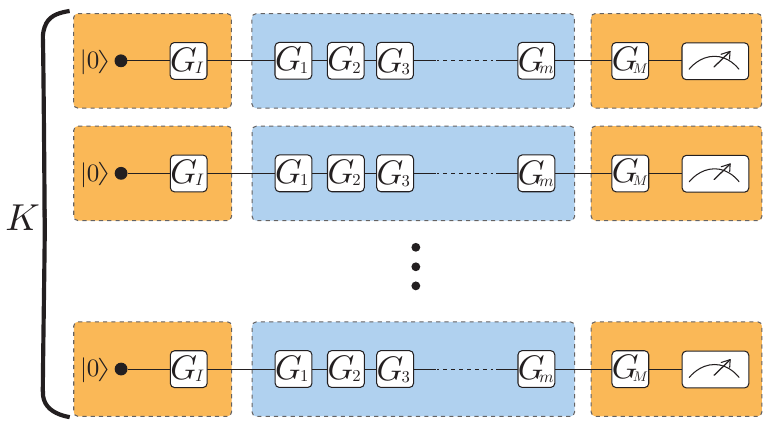}
	\caption{A DRB circuit in the single-qubit case is composed as follows. The processes $G_I$ and $G_M$ correspond to stabilizer state preparation and stabilizer measurement, respectively. The superoperators $G_1, G_2, \ldots, G_m$ represent randomly generated generators of the Clifford group. $K$ represents the number of randomized circuits with depyh $m$.}
	\label{fig:rb_circs}
\end{figure}

A DRB circuit begins with stabilizer state preparation \cite{ref:st_states_1, ref:st_states_2} from \( \ket{0} \) via the noisy initialization operation \( \tilde{G}_I \), followed by \( m \) randomly selected layers of noisy quantum processes \( \tilde{G}_1, \dots, \tilde{G}_m \), and ends with the application of \( \tilde{G}_M \) prior to measurement (see Fig.~\ref{fig:rb_circs}). Each component introduces a small deviation from the intended evolution due to experimental noise. An example gate set is \( \{ R_x(\pm \pi/2), R_y(\pm \pi/2), R_x(\pm \pi), R_y(\pm \pi) \} \).

In the single-qubit case, Pauli group stabilizer states are eigenvectors of the Pauli operators. They can be prepared using a single gate from the set $\{ I,\, R_y(\pm \pi/2),\, R_y(\pi),\, R_x(\pm \pi/2) \},$ producing six distinct stabilizer states.

In the superoperator formalism, the probability of obtaining the target outcome is written as:
\begin{equation}
	P = \vec{M}_{\text{target}}^{\dagger} \tilde{G}_M \tilde{G}_C \tilde{G}_I \vec{\rho}_{\text{in}},
\end{equation}
where \( \tilde{G}_C \) denotes the noisy implementation of the core sequence, and \( \vec{M}_{\text{target}} \) is the POVM element corresponding to the desired outcome.

Empirically, this success probability is estimated by:
\begin{equation}
    P = \frac{N_{\text{target}}}{N},
\end{equation}
with \( N \) the number of measurement repetitions, and \( N_{\text{target}} \) the count of outcomes matching the target state.

The observed success probability for generated circuits decays exponentially with increasing circuit depth. DRB theory approximates this decay by the formula \cite{ref:RB_5}:
\begin{equation}
	P = A p^{m} + B,
\label{eq:succ_prob}
\end{equation}
where \( A \) and \( B \) reflect SPAM contributions and \( p \) is the depolarization parameter. The average gate fidelity is then extracted as \cite{ref:RB_3}:

\begin{equation}
	F^{avg}_{RB} = p + \frac{1 - p}{d},
\label{eq:avg_fid_rb}
\end{equation}
where $d = 2$ denotes the dimension of the Hilbert space.

Thus, using equations~\eqref{eq:succ_prob} and \eqref{eq:avg_fid_rb}, gate operation accuracy metrics can be directly derived.

\subsection{Gate set tomography}

Gate set tomography (GST) is a self-consistent tomographic protocol that reconstructs not only the quantum gates under study but also the state preparation and measurement operations~\cite{ref:GST_1, ref:GST_2, ref:GST_3}. Unlike standard process tomography, GST does not assume ideal preparation and measurement and thus eliminates their contribution to systematic errors.

\begin{figure}[t!]
	\centering
	\includegraphics[width=1.0\linewidth]{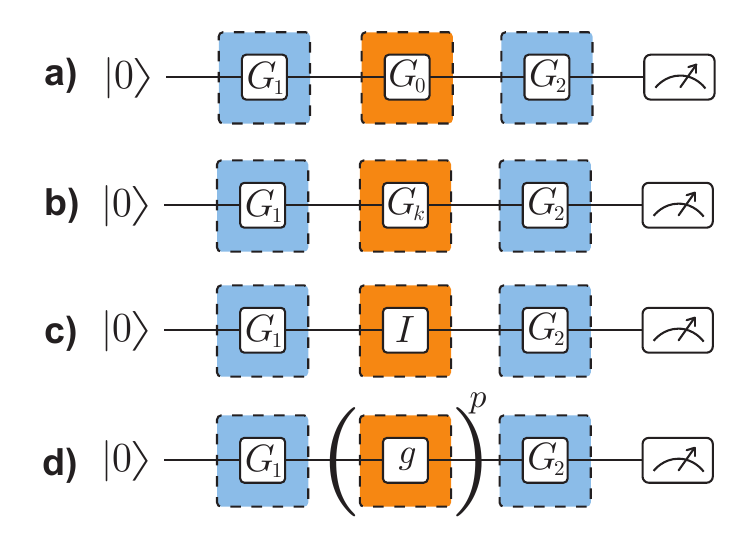}
	\caption{(\textbf{a})–(\textbf{c}) GST circuits. The quantum process under investigation is denoted as $G_0$, while $G_1$ and $G_2$ represent the superoperators corresponding to state preparation and measurement, respectively. $G_1$ and $G_2$ are incorporated into the circuits as $G_k$. In the final set of circuits, the process under investigation is the identity (trivial) operation.
	\textbf{(d)} Additional circuits used in long-sequence GST. Here, $g$ denotes a germ, which is a sequence composed of gates from the gate set, and $p$ corresponds to the number of repetitions of the germ $g$ within the circuit.}
	\label{fig:gst}
\end{figure}

GST circuits consist of sequences where a gate of interest is inserted between fixed preparation and measurement operations (see Fig.~\ref{fig:gst}a). To achieve full reconstruction, circuits also include gates for preparation and measurement (see Fig.~\ref{fig:gst}b) and identity operation (see Fig.~\ref{fig:gst}c), enabling a form of linear inversion. To increase precision, GST uses so-called germs -- short sequences of gates repeated multiple times (up to $L$ times), which amplify systematic errors and help identify gate-specific imperfections (see Fig.~\ref{fig:gst}d).

Linear inversion provides an initial estimate of the gate set by solving:
\begin{equation}
    p_{kij} = \vec{M}_i^{\dagger} G_k \vec{\rho}_j,    
\end{equation}
where the vectors represent the measurement operators and input states, and \( G_k \) are the superoperators to be reconstructed. The linear inversion algorithm is described in details in \cite{ref:GST_2} and in Appendix~\ref{gopt}1. 

The reconstruction is refined using maximum likelihood estimation (MLE), maximizing the log-likelihood function:
\begin{equation}
    \ln(L) = \sum_{m,\alpha_m} N_m \ p^{(m,\alpha_m)}_{\mathrm{exp}} \ln \left( p^{(m,\alpha_m)}_{\mathrm{model}} \right),
\end{equation}
where $m$ indexes the generated circuits, $\alpha_m$ denotes the measurement outcome for circuit $m$, and $N_m$ is the number of repetitions. The observed frequencies are given by \( p^{(m, \alpha_m)}_{\text{exp}} = N_{m, \alpha_m}/N_m \), with $N_{m, \alpha_m}$ being the number of times outcome $\alpha_m$ is observed. The values \( p^{(m, \alpha_m)}_{\text{model}} \) are the corresponding predictions from the reconstructed model. Results from linear inversion provide a solid starting point for maximum likelihood estimation optimization.

To ensure that physical constraints such as complete positivity and trace preservation are satisfied, the optimization is carried out using Riemannian geometry on the Stiefel manifold~\cite{luchnikov2021riemannian}. This method guarantees that the resulting superoperators remain within the physical space of quantum processes. 

The reconstruction procedure in GST is subject to gauge freedom, as different sets of quantum operations -- potentially corresponding to slightly different physical processes -- can produce identical measurement statistics. As a result, quantities such as fidelity, which are not gauge-invariant, cannot be uniquely determined without additional assumptions. To address this ambiguity, gauge optimization is typically performed by aligning the reconstructed processes with a target set of ideal operations. One approach based on convergence toward the expected ideal form is described in~\cite{ref:GST_1} and in Appendix~\ref{gopt}2.

\section{\label{sec:simulation}Simulation of benchmarking algorithms}

In this section, we present simulation results obtained using the DRB and GST protocols described above. A numerical simulator was developed to model realistic quantum channels, incorporating both unitary gate operations and specific noise processes. The noise model includes longitudinal (\( T_1 \)) and transverse (\( T_2 \)) relaxation times, along with SPAM errors characterized by transition probabilities \( p_{0\rightarrow1} \) and \( p_{1\rightarrow0} \) for incorrect measurement outcomes.

\begin{figure}[t!]
	\centering
	\includegraphics[width=1.0\linewidth]{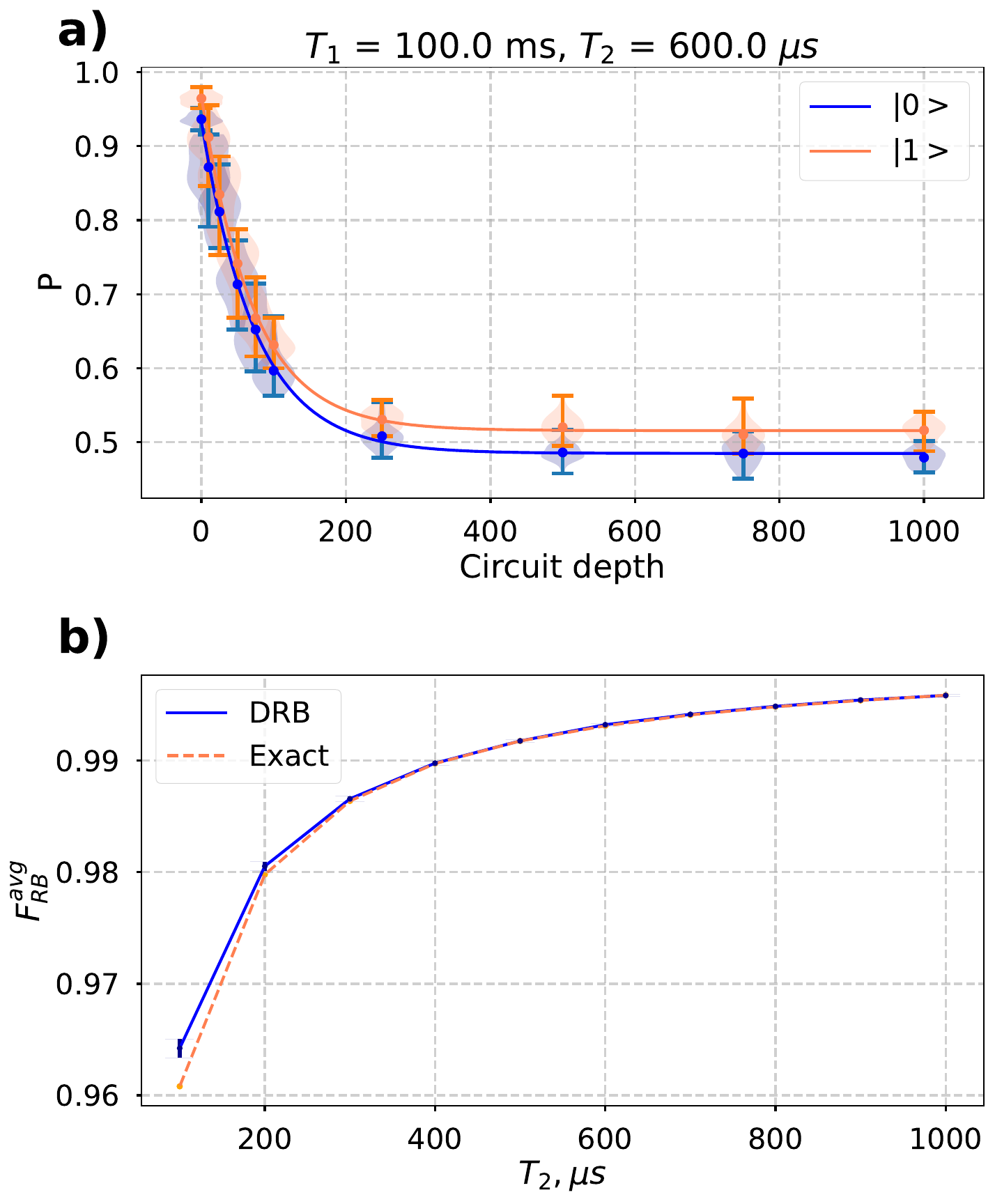}
        \caption{(a) Simulated success probability \( P \) as a function of circuit depth \( m \) for \( T_2 = 600~\mu\text{s} \). Results for outcomes \( \ket{0} \) and \( \ket{1} \) are shown separately to highlight the impact of SPAM asymmetry. The shaded region indicates statistical spread across all circuits. Error bars demonstrate only the size of these regions.\\
        (b) Average gate fidelity \( F^{\text{avg}}_{RB} \) estimated from DRB simulation (solid line with $95\%$ confidence interval error bars) compared to theoretical predictions based on superoperator analysis (dashed line). Good agreement is observed, with small deviations for short \( T_2 \) reflecting the breakdown of DRB assumptions in high-noise regimes.}
	\label{fig:1q_sim_drb}
\end{figure}

\subsection{Direct Randomized Benchmarking Simulations}

Simulations were performed for single-qubit circuits using the DRB protocol. The input state was \( \rho = \dyad{0} \), and measurements were carried out in the computational basis. Each circuit was repeated \( N = 1000 \) times to estimate probabilities. The gate set included single-qubit rotations \( R_x(\theta) \) and \( R_y(\theta) \) with \( \theta = 0, \pm\pi/2, \pi \), which suffices to generate the Clifford group.

We simulated circuits with depths \( m = 0, 25, 50, 100, 250, 500, 750, 1000 \), using \( K = 25 \) randomized circuits per depth targeting the outcome \( \ket{0} \), and another 25 targeting the outcome \( \ket{1} \). The success probability \( P(m) \) was fitted using Eq.~\eqref{eq:succ_prob} to extract the depolarization parameter \( p \), and the average fidelity was then computed via equation~\eqref{eq:avg_fid_rb}.

The primary objective of this simulation study was to assess the applicability of the benchmarking protocol to experimentally feasible gate operations and to validate its performance on a neutral atom processor. To that end, we focused on single-qubit systems.

The simulation was performed with the following parameters: $T_1 = 100$ ms that is close to the physical quantum computer, $T_2$ is varied from $100$ to $1000$ $\mu$s, $p_{0\rightarrow1} = 6\%$, $p_{1\rightarrow0} = 3\%$. These values were chosen to be consistent with the experimentally measured asymmetric readout errors of our setup.

Fig.~\ref{fig:1q_sim_drb}a illustrates the decay of success probability for a fixed \( T_2 = 600~\mu\text{s} \). The difference between outcomes \( |0\rangle \) and \( |1\rangle \) reflects asymmetry in SPAM errors. The shaded regions in the violin plot represent the distribution of success probabilities across all $K$ circuits of a given depth. In other words, they correspond to the individual success probabilities obtained for each of the $K$ circuits. The overlaid curves, in turn, show the average trend of success probability across all circuits at each depth. In Fig.~\ref{fig:1q_sim_drb}b, the average fidelity obtained from DRB closely follows the theoretical fidelity calculated using the explicit error superoperator. This validates both the simulator and the fidelity estimation procedure. 
Minor discrepancies at short $T_2$ times reflect stronger error regimes that marginally exceed the descriptive scope of DRB.

By fitting DRB decay curves separately for circuits targeting the outcomes \(|0\rangle\) and \(|1\rangle\), we extract distinct parameters \(A_0, B_0, p_0\) and \(A_1, B_1, p_1\), respectively. For each case, the total SPAM error can be estimated as \(p_{0 \rightarrow 1} = 1 - (A_0 + B_0)\) and \(p_{1 \rightarrow 0} = 1 - (A_1 + B_1)\). For \(T_2 = 600\,\mu\text{s}\), we obtain \(p_{0 \rightarrow 1} = 6.38^{+0.27}_{-0.25}\%\) and \(p_{1 \rightarrow 0} = 3.37^{+0.22}_{-0.25}\%\), both reported with a 95\% confidence interval (see. Appendix~\ref{app:bootstrap}).

\begin{figure*}[t!]
	\centering
	\includegraphics[width=1\linewidth]{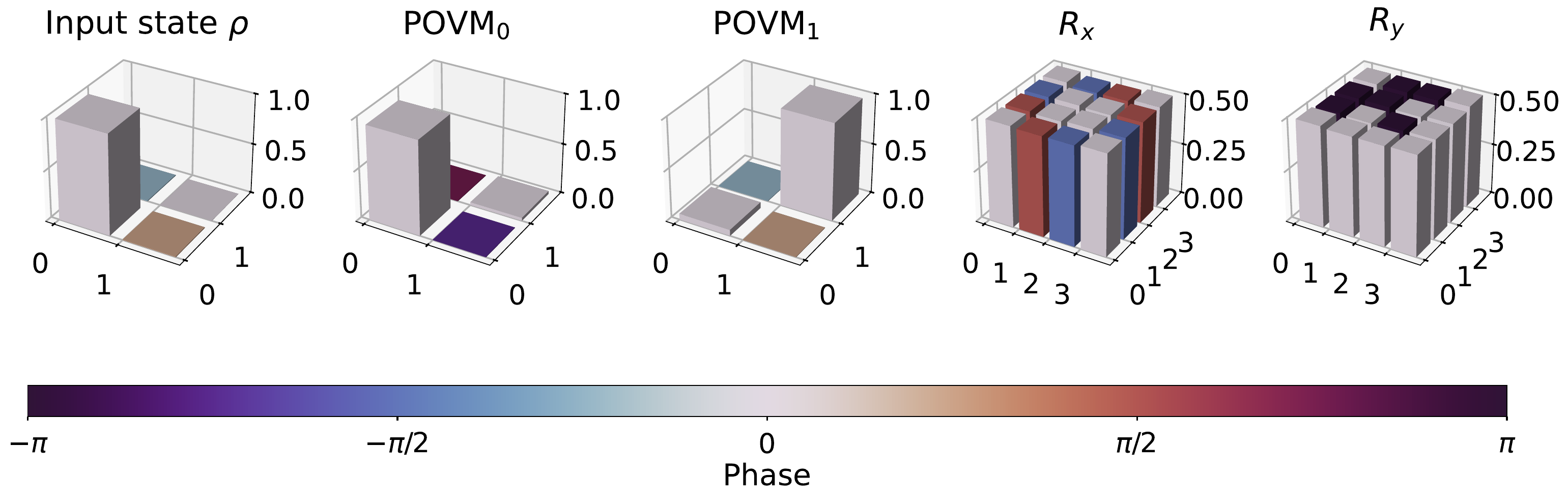}
	\caption{Reconstructed superoperators for GST at \( T_1 = 100~\text{ms}, T_2 = 600~\mu\text{s} \), and SPAM probabilities \( p_{0\rightarrow1} = 1 \% \), \( p_{1\rightarrow0} = 25 \% \). The diagrams show: (top center) input state; (top left, top right) POVM elements for \( \ket{0} \) and \( \ket{1} \); (bottom left, bottom right) superoperators for \( R_x(\pi/2) \) and \( R_y(\pi/2) \). The numerical reconstructions match the parameters of the simulation with high fidelity.}
	\label{fig:1q_sim_gst_tomo}
\end{figure*}

\subsection{Gate Set Tomography Simulations}

Simulations were also conducted for GST using the same error model. Circuits were constructed using \( R_x(\pi/2) \) and \( R_y(\pi/2) \) gates, with germs \( R_xR_y \) and \( R_yR_x \), each repeated up to three times (\( L = 3 \)). Each circuit was repeated \( N = 1000 \) times to estimate probabilities. The input state was \( \rho = \dyad{0} \), and measurements were carried out in the computational basis. Linear inversion was used to initialize MLE optimization. 

The GST protocol enables the extraction of significantly more information about quantum processes than DRB, as it fully reconstructs the matrices of all superoperators. Additionally, it provides estimates for the input state and measurement operators.

Fig.~\ref{fig:1q_sim_gst_tomo} presents GST results for the same noise parameters. The reconstructed operators yield the following estimates: the readout error probabilities are $p_{0\rightarrow1} = 5.97^{+0.35}_{-0.34}\%$, $p_{1\rightarrow0} = 3.26^{+0.35}_{-0.35}\%$, all reported with $95\%$ confidence interval. Gate fidelities were computed using equation~\eqref{eq:fid_a_e_couple} based on entanglement fidelity extracted from the reconstructed superoperators.

Error bars indicate $95\%$ confidence intervals obtained via a nonparametric bootstrap procedure. Simulated frequencies were treated as ground-truth values to generate new synthetic datasets by sampling from a binomial distribution. The benchmarking algorithm was applied identically to each resampled dataset. The standard deviation of the resulting estimates was used to construct the confidence intervals.

\begin{figure}[b!]
	\centering
	\includegraphics[width=1.0\linewidth]{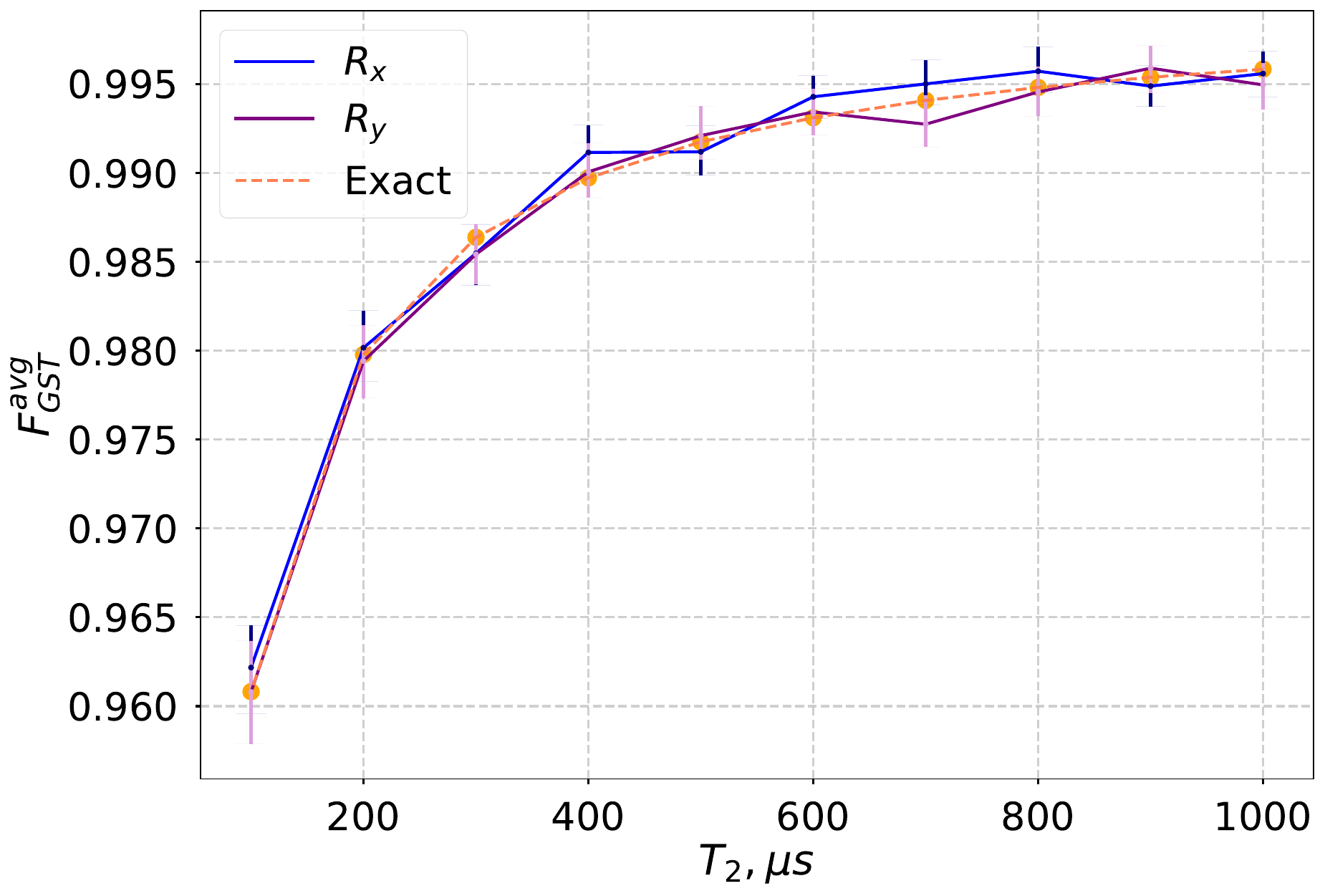}
	\caption{Fidelity of \( R_x \) and \( R_y \) gates reconstructed by GST as a function of transverse relaxation time \( T_2 \). Simulation results (solid lines with $95\%$ confidence interval error bars) are compared with theoretical predictions (dashed lines). The agreement across a wide range of \( T_2 \) values confirms the accuracy of GST reconstruction.
    }
	\label{fig:1q_sim_gst_fid}
\end{figure}

The dependence of gate fidelity on \( T_2 \) is shown in Fig.~\ref{fig:1q_sim_gst_fid}. Here, our proposed gauge-fixing method is applied; a detailed description of this procedure is provided in Appendix~\ref{gopt}.

Overall, the DRB and GST protocols demonstrate strong agreement with theoretical predictions and with each other. While DRB offers a fast and SPAM-robust estimate of average fidelity, GST provides a more detailed picture of quantum operations at the cost of increased experimental and computational resources.

\section{\label{sec:experiment}Benchmarking of a physical device}

We now present experimental benchmarking results obtained on a neutral atom quantum processor. The processor is based on an array of single rubidium atoms in tightly focused optical tweezers. \(^{87}\text{Rb}\) atoms are cooled in a magneto-optical trap (MOT) and loaded into a tweezer array with a final temperature of $\sim$50~$\mu$K. We use an array of tweezers which is holographically generated with far-off-resonant light at the wavelength of 813~nm passing through a reflective liquid crystal spatial light modulator. The tweezers are tightly focused by a 0.6 NA aspherical lens installed inside a vacuum chamber to a beam waist of $\sim1$~$\mu$m. The distance between the tweezers in the array is 3.5~$\mu$m, and individual sites are well resolved with an imaging system based on a second in-vacuum aspheric lens and additional optics that image the atomic fluorescence on a CMOS camera. To minimize the variance of the trap depths we adjust the holograms using an iterative procedure based on the weighted Gerchberg-Saxton algorithm with the weights adjusted to minimize the difference in the measured atomic fluorescence signal \cite{ref:T_16}. This procedure allows us to achieve the variance of fluorescence signal below $2.5\%$.

Qubits are encoded in hyperfine sublevels of the ground state of \(^{87}\text{Rb}\), specifically \( \ket{0} = \ket{5S_{1/2}, F=2, m_F=0} \) and \( \ket{1} = \ket{5S_{1/2}, F=1, m_F=0} \). Initially, the atoms are optically pumped to the $\ket{0}$ state, then single-qubit gates are implemented via microwave-driven transitions using a global RF field. The microwave frequency is adjusted to $6834.686$ MHz in the Ramsey-like experiments. This corresponds to a +4 kHz shift from the hyperfine clock transition frequency in vacuum due to the quadratic Zeeman shift in the bias magnetic field of 4.3 Gauss and the differential AC Stark shift from the tweezer laser which is on during single qubit operations. Site-resolved measurements are realized using a $\sigma^+$-polarized push-out beam resonant with the $\ket{5S_{1/2}, F=2} \rightarrow \ket{5\,P_{3/2}, F=3}$ transition that expels the atoms in the qubit state $\ket{0}$ from the trap, followed by fluorescence imaging with MOT light. The push-out beam is shaped with a spatial light modulator and focused with the same high-NA lens to a pattern exactly matching the tweezer array.

Experimental value of $T_1$ time is about 100~ms, $T_2^*$ time is about 1400~$\mu$s, the latter is measured in a Ramsey-like experiment, and $T_2$ time of 8~ms is measured in a standard echo experiment.

We apply the DRB protocol in two experimental settings — a single isolated qubit and a 25-qubit two-dimensional array — to assess individual gate fidelities and the spatial uniformity of global control. Complementary GST measurements were performed on a separate isolated qubit to provide detailed tomographic verification. These complementary scenarios allow us to characterize both individual gate performance and the spatial homogeneity of global control in a large-scale system. Our results demonstrate high single-qubit fidelities, effective correction of coherent errors via DRB-based calibration, and robust spatial performance across a 25-qubit array.

\begin{figure}[t!]
	\centering
	\includegraphics[width=1.0\linewidth]{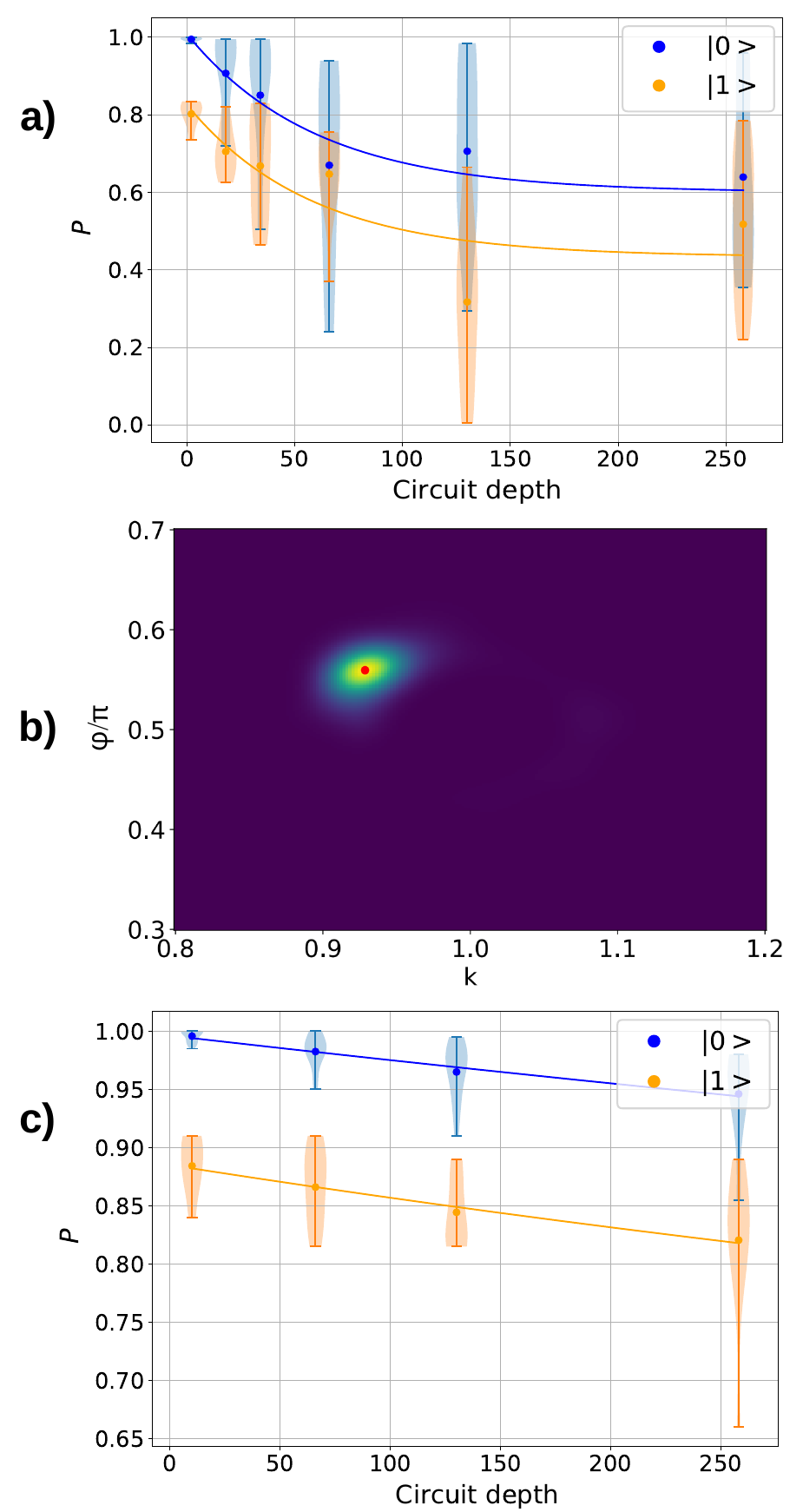}
    \caption{(a) DRB success probabilities before calibration, shown separately for outcomes \( \ket{0} \) and \( \ket{1} \). Shaded violin plots show the distribution of success probabilities across $K$ circuits per depth, revealing variability due to circuit-specific noise effects. Solid curves represent the average trend across all circuits at each depth. A significant fraction of circuits exhibit success probabilities below 0.5, indicating output states orthogonal to the target.\\
    (b) Calibration map derived from simulated DRB results for a grid of control parameters. The experimental setting lies outside the optimal region, which reveals the presence of systematic gate miscalibration. The values on the vertical axis are represented in the units of $\pi$ angle.\\
    (c) DRB performance after applying the calibration correction. Success probabilities are markedly improved, with fidelity values consistent across outcomes.
    }
	\label{fig:drb_planes}
\end{figure}

\begin{figure*}[t!]
	\centering
	\includegraphics[width=1.0\linewidth]{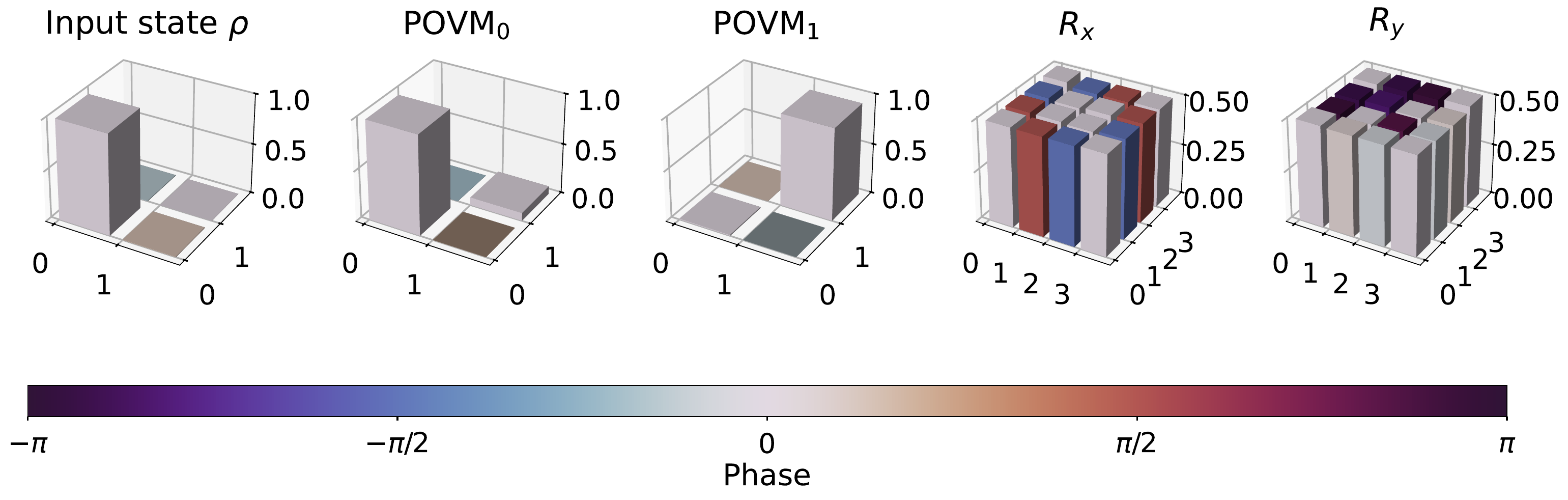}
        \caption{Reconstructed superoperators for GST. The diagrams show: (top center) input state; (top left, top right) POVM elements for \( \ket{0} \) and \( \ket{1} \); (bottom left, bottom right) superoperators for \( Rx(\pi/2) \) and \( Ry(\pi/2) \).
    }
	\label{fig:tomo_pictures}
\end{figure*}

\subsection{Single-Qubit Experiment}

We first performed DRB benchmarking on a single qubit. Circuits were generated with depths \( m = 0, 16, 32, 64, 128, 256 \), 
using \( K = 10 \) randomized circuits per depth targeting the outcome 
\( \ket{0} \), and another $10$ targeting the outcome \( \ket{1} \). 
Each circuit was executed \( N = 200 \) times. The gate set included \( R_x(\theta), R_y(\theta) \) with \( \theta = 0, \pm \pi/2, \pm\pi \), which suffices for generating all single-qubit Clifford operations.

As shown in Fig.~\ref{fig:drb_planes}a, the pre-calibration fidelity curve exhibits pronounced variations between \( \ket{0} \) and \( \ket{1} \) outcomes. The shaded regions in the violin plot have the same interpretation as described previously, representing the distribution of success probabilities over the set of $K$ circuits at each depth. The overlaid curves indicate the corresponding average trend. The average gate fidelity from these curves is $F^{avg}_{RB} = 99.360 ^{+0.057}_{-0.059}\%$ with a $95\%$ confidence level.

Notably, the wide spread of success probabilities across circuits of the same depth suggests that noise effects may vary significantly with circuit structure. This behavior is indicative of nontrivial coherent or gate-dependent error mechanisms. In particular, a considerable number of sequences yield success probabilities below 0.5, corresponding to inverted measurement outcomes -- an outcome incompatible with the high average fidelity extracted from fitting. This discrepancy strongly suggests the presence of coherent calibration errors in the gate implementation.


To correct for systematic control errors, we developed a two-parameter calibration model integrated with the DRB protocol and applied iteratively: DRB is first performed to extract calibration parameters, followed by remeasurement with updated control settings. This model captures coherent deviations in gate implementation through two physically motivated parameters: an overrotation factor $k$, arising from imperfect calibration of the $\pi$-pulse duration of the laser, and an azimuthal offset angle $\varphi$, which accounts for phase misalignment between the nominal $R_x$ and $R_y$ gates. In practice, the $x$-axis orientation is fixed by definition, while the correct $y$-axis orientation is realized by adding a phase shift to the laser drive, directly corresponding to the parameter $\varphi$.

The calibration parameters are extracted by minimizing the deviation between simulated and experimentally measured circuit outcomes across the entire DRB dataset, without relying on auxiliary calibration routines or prior assumptions. The procedure is detailed in Appendix~\ref{calibr}, and the extracted values of \( k \) and \( \varphi \) are visualized in the calibration map shown in Fig.~\ref{fig:drb_planes}b.

After applying the correction, a new set of DRB circuits was generated and measured with the updated control parameters. The resulting DRB curves exhibit a substantial improvement, as seen in Fig.~\ref{fig:drb_planes}c. The recalibrated average gate fidelity is $F^{\text{avg}}_{\text{RB}} = 99.963 ^{+0.015}_{-0.013}\%$ with a $95\%$ confidence level, representing a significant improvement over the pre-corrected value.

By fitting the decay curves to the model of equation~\eqref{eq:succ_prob}, we extract parameters $A$ and $B$, yielding $p_{0\rightarrow1} = 1.49 ^{+0.41}_{-0.38}\%$ and $p_{1\rightarrow0} = 11.86 ^{+1.31}_{-1.28}\%$, both reported with a $95\%$ confidence interval. The pronounced difference in state measurement efficiency observed in the experiments is explained by the asymmetry in the implemented measurement procedure. Our experiment uses a resonant ejection laser beam, which pushes out the atoms in the quantum state $\ket{0}$ only. However, an atom can be lost during the experiment due to finite lifetime in the trap or other factors unrelated to its quantum state. In the push-out measurement these losses will be interpreted as an outcome corresponding to the quantum state $\ket{0}$, significantly increasing the SPAM error for the state $\ket{1}$. SPAM error may be reduced by careful optimization of the push-out beam parameters. However, for the purposes of this study it is interesting to note, that the DRB protocol and the calibration procedure tolerate significant asymmetry in the SPAM errors.

After the DRB we performed another experiment with GST measurements. Long-sequence circuits were constructed with repetition length \( L = 3 \), and each was executed \( N = 200 \) times. Fig.~\ref{fig:tomo_pictures} shows the reconstructed tomographic data. The input state fidelity is \( 99.98 ^{+0.04}_{-0.05}\% \), indicating near-ideal preparation. Measurement errors were estimated as \( p_{0\rightarrow1} = 5.59 ^{+0.25}_{-0.24}\% \), \( p_{1\rightarrow0} = 2.56 ^{+0.16}_{-0.15}\% \). The reconstructed gate fidelities are \( F^{avg}_{x, \text{GST}} = 99.94 ^{+0.02}_{-0.11} \% \) for \( R_x \) and \( F^{avg}_{y, \text{GST}} = 99.38 ^{+0.13}_{-0.20}\% \) for \( R_y \), consistent with the DRB results. Each value is reported with a $95\%$ confidence interval. The observed discrepancy between the two gates reflects coherent miscalibration errors affecting the $R_y$ implementation.

\subsection{Experiment with qubit array}

Finally, we applied the DRB protocol to a 25-qubit register with global single-qubit control. Benchmarking was performed individually for each atom in the array using single-qubit circuits. While an overall trend in fidelity was observed, some qubits showed statistically significant deviations from the mean.

\begin{figure}[t!]
    \centering
    \includegraphics[width=1\linewidth]{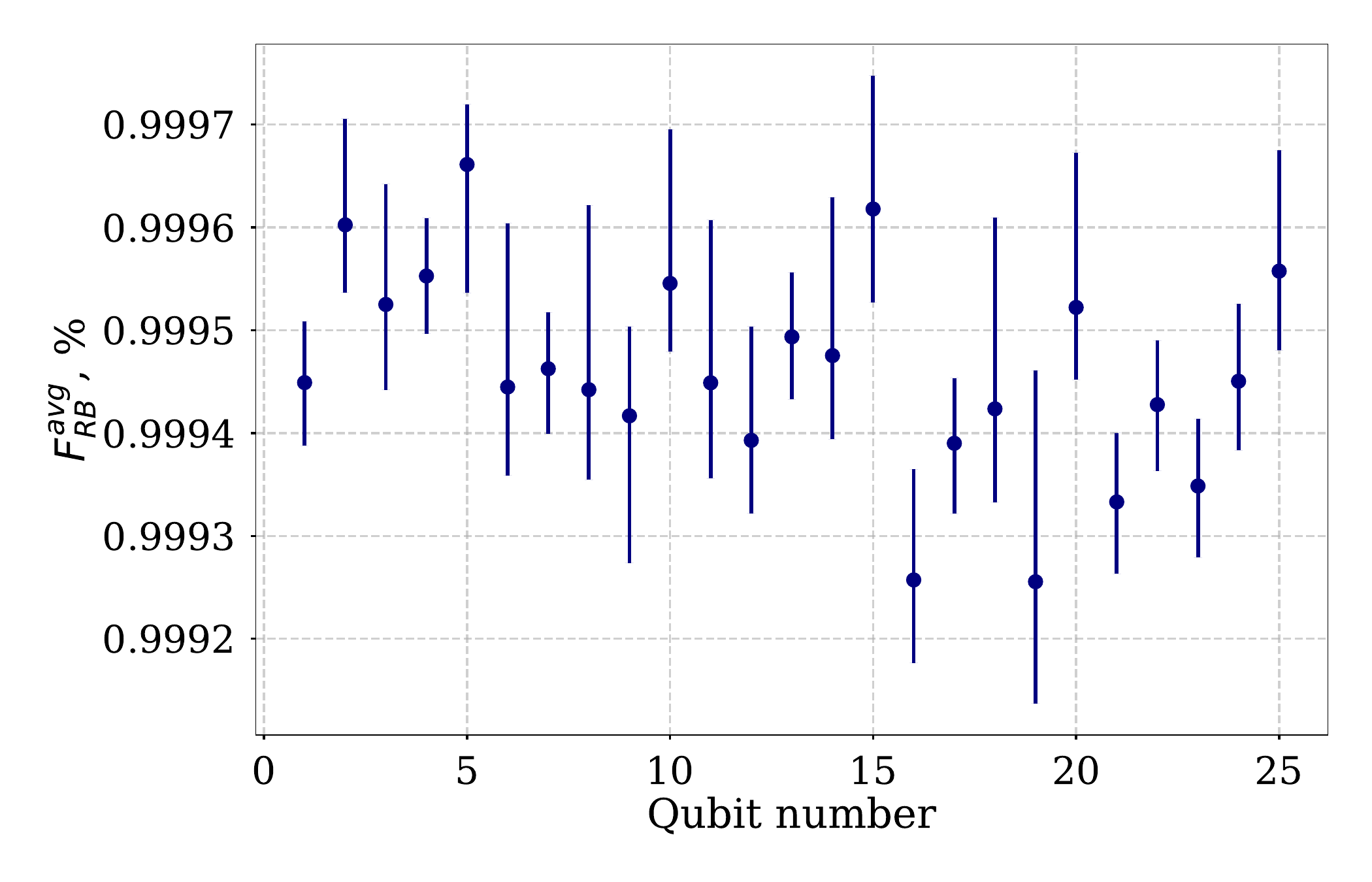}
        \caption{Estimated gate fidelity \( F^{\text{avg}}_{\text{RB}} \) for each qubit in a 25-qubit array. Qubit performance varies significantly across sites, with statistically distinguishable differences observed between certain locations.
    }
    \label{fig:25_q}
\end{figure}

As shown in Fig.~\ref{fig:25_q}, spatially dependent variations in gate fidelity reflect the underlying inhomogeneity of the system. The average fidelity across all 25 sites was \( \langle F^{\text{avg}}_{\text{RB}} \rangle = 99.946 ^{+0.002}_{-0.001}\% \), confirming that global control achieves nearly the same fidelity as isolated control, with inhomogeneities remaining within acceptable limits.

\section{Conclusion}

In this work, we performed a comparative study of two quantum benchmarking techniques -- Direct Randomized Benchmarking (DRB) and Gate Set Tomography (GST) -- on both simulated and experimental single-qubit systems based on neutral atom technology. Using a custom numerical simulator incorporating realistic relaxation and SPAM noise models, we validated the agreement between theoretical predictions and the outputs of both benchmarking protocols. Our analysis confirmed the ability of DRB to yield rapid and SPAM-robust fidelity estimates, while GST provided detailed tomographic insight into gate implementations.

In addition to simulations, we applied these methods to a neutral atom processor. On a single qubit, DRB revealed coherent control errors, which we corrected via a two-parameter calibration procedure. The corrected gates achieved an average fidelity of $99.963 ^{+0.015}_{-0.013}\%$. GST measurements on the same system independently verified high-quality state preparation and gate fidelity.

Applying the DRB protocol across a 25-qubit array, we benchmarked each qubit individually under global control fields. The resulting average fidelity of \( 99.946 ^{+0.002}_{-0.001}\% \) remained consistent with the single-qubit result, despite spatial variations across the array.

We also introduced a novel gauge-fixing approach for GST analysis, involving the diagonalization of state preparation and measurement operators followed by fidelity-based optimization of gate orientation. This procedure improves the interpretability of reconstructed gates and enables meaningful fidelity comparisons.

Collectively, our results establish the viability of scalable and precise single-qubit control in neutral atom systems and demonstrate the utility of employing complementary benchmarking protocols to probe quantum gate performance under diverse conditions.

\begin{acknowledgments}
Theoretical part of this work was supported by Russian Science Foundation, project 22-12-00263-$\Pi$, \url{https://rscf.ru/en/project/22-12-00263/}. The authors acknowledge support by Rosatom in the framework of the Roadmap for Quantum computing (Contract No.P2154 dated November 24, 2021).
\end{acknowledgments}

\section*{Data availability}

The data supporting this study are available in Zenodo; see Ref.~\cite{ref:bench_dataset}.

\bibliography{bible.bib}

\appendix
\section{\label{gopt}Gauge optimization procedure}

\subsection*{A1. Linear inversion and calibration initialization}

As an initial step in GST, we employ a linear inversion method to obtain first-order estimates of the quantum processes, input states, and measurements. The idealized probability of a measurement outcome in a GST experiment is given by
\begin{equation}
    p_{kij} = \vec{M}_i^{\dagger} G_k \vec{\rho}_j,
\end{equation}
where \( \rho_j \) and \( M_i \) are the prepared state and POVM element, respectively, implemented via gate sequences represented by quantum processes \( F_j \) and \( F_i \). Assuming \( \rho_0 \) is the default prepared state and \( M_0 \) is the native measurement operator (typically in the computational basis), the probability can be rewritten as
\begin{equation}
    p_{kij} = \vec{M}_0^{\dagger} F_i G_k F_j \vec{\rho}_0.
\end{equation}
    
To solve for \( G_k \), we expand the space of matrices using a basis \( C_{mn} = \ket{m}\bra{n} \), where \( m, n = 0, 1 \) (in single-qubit case), and define their vectorized forms \( \vec{C}_\alpha \) such that
\begin{equation}
    \sum_\alpha \vec{C}_\alpha \vec{C}_\alpha^\dagger = I.
\end{equation}

By inserting this decomposition into the probability expression, the model can be cast into matrix form:
\begin{equation}
    p_k = A G_k B,
\end{equation}
where the matrix elements are defined as

\begin{equation}
    A_{i\alpha} = \vec{M}_0^{\dagger} F_i \vec{C}_\alpha, \quad B_{\beta j} = \vec{C}_\beta^\dagger F_j \vec{\rho}_0.
\end{equation}

We define the Gram matrix as
\begin{equation}
    g = A B,
\end{equation}
which, from its construction, corresponds to a matrix of probabilities obtained from experimental data using the identity process \( G_k = I \). For this reason, all GST gate sequences must include a configuration with the identity operation as the target process.

It follows that each process \( G_k \) can be reconstructed as
\begin{equation}
    G_k = B g^{-1} p_k B^{-1}.
\end{equation}
    
Let \( \vec{R} \) denote the first column of \( g \) and \( \vec{Q}^T \) its first row. Then, the reconstructed state and POVM elements are obtained via
\begin{equation}
    \vec{\rho} = B g^{-1} \vec{R}, \quad \vec{E}^T = \vec{Q}^T B^{-1}.
\end{equation}

The matrix \( B \), however, depends on the actual (potentially noisy) implementation of the gate set \( F \), which is not directly known. To proceed, we initialize the calibration by choosing \( B \) as if the state preparation and measurement were ideal. This choice defines a gauge frame and serves as the starting point for the subsequent optimization.

\subsection*{A2. Post-MLE gauge fixing}

Once MLE is applied, the reconstructed objects no longer preserve the gauge frame defined by the idealized \( B \), due to numerical fluctuations and statistical noise. To resolve this ambiguity and place the reconstructed gates in a physically meaningful frame, we perform a gauge optimization.

First, we apply a joint diagonalization procedure to find a unitary transformation \( G_U \) that approximately diagonalizes both the reconstructed input state and POVM elements. This operation identifies a common eigenbasis and aligns them to canonical axes. The optimization is carried out by minimizing the off-diagonal weight function:
\begin{equation}
f(G_U) = \sum_{\substack{i,j \\ i \neq j}} \left| (G_U^\dagger \rho G_U)_{i,j} \right|^2 + \sum_{\substack{i,j,k \\ i \neq j}} \left| (G_U^\dagger O_k G_U)_{i,j} \right|^2
\label{eq:diag_func}
\end{equation}
where sum runs over the reconstructed state $\rho$ and POVM operators $O_k$.

Next, we apply an additional gauge rotation \( R_z(\delta) \), where the gauge angle \( \delta \) is selected to bring the reconstructed \( R_x \) and \( R_y \) gates into canonical form. The value of \( \delta \) is determined by maximizing the average fidelity between the transformed superoperators and their ideal targets:
\begin{equation}
F(\delta) = \frac{\mathrm{Tr}\left[G_{\mathrm{id}}^{\dagger} G^{\dagger}_{R_z}(\delta) \tilde{G} G_{R_z}(\delta)\right]}{6} + \frac{1}{3},
\end{equation}
where $\tilde{G}$ denotes the GST-estimated superoperator transformed by $G_{R_z}(\delta)$, and $G_{id}$ is the corresponding ideal target.

\section{\label{calibr}Calibration model based on coherent error parameters}

To account for systematic coherent errors in gate implementations, we employ a two-parameter correction model. Specifically, we assume that each single-qubit gate deviates from its ideal form by an overrotation factor \( k \) and an azimuthal angle offset \( \varphi \), such that:
\begin{equation}
R_x(\theta) \rightarrow R_x(k\theta), \quad R_y(\theta) \rightarrow R_\varphi(k\theta),
\end{equation}
where \( R_\varphi(k\theta) \) denotes a rotation by angle \( k\theta \) about an axis in the \( xy \)-plane rotated by angle \( \varphi \) from the \( x \)-axis. These two parameters are chosen to describe the dominant coherent errors observed in the neutral atom platform and have a direct physical motivation. The amplitude scaling factor $k$ arises from imperfect calibration of the laser $\pi$-pulse duration: if the pulse is slightly too short or too long, the resulting gate implements an over- or under-rotation. The axis shift $\varphi$, on the other hand, captures a phase misalignment between nominally orthogonal $R_x$ and 
$R_y$ gates. In our implementation, the x-axis is fixed by definition, while the correct y-axis operation corresponds to a phase shift of $\pi/2$ in the laser drive. Deviations from this ideal phase are therefore naturally parameterized by $\varphi$.

The true gate parameters are extracted by minimizing the mismatch between experimentally measured success probabilities and simulations generated using the model above. For each randomized benchmarking circuit, we define an individual error function:
\begin{equation}
f_i = \left( P_\text{sim}^{(i)}(k, \varphi) - P_\text{exp}^{(i)} \right)^2,    
\end{equation}
and construct a global objective function:
\begin{equation}
f_{\times}(k, \varphi) = \prod_i \left( 1 - f_i \right),
\end{equation}
where the product runs over all benchmarking circuits in the dataset. The optimal calibration parameters correspond to the maximum of \( f_{\times}(k, \varphi) \), as illustrated in Fig.~\ref{fig:drb_planes}b.

Once optimal values are determined, the corrected experimental control parameters are updated via:
\begin{equation}
\tau_\text{correct} = \frac{\tilde{\tau}}{k}, \quad \varphi_\text{correct} = \tilde{\varphi} - (\varphi - \pi/2),
\end{equation}
where \( \tilde{\tau} \) and \( \tilde{\varphi} \) denote the uncalibrated pulse duration and azimuthal angle, respectively. This procedure ensures that control errors are compensated at the level of physical pulse parameters.

\section{\label{app:bootstrap}Uncertainty Estimation via Bootstrap}

To estimate uncertainties in gate fidelity and other bounded metrics, we employ the percentile bootstrap method~\cite{ref:bootstrap}. This non-parametric approach makes no distributional assumptions and naturally respects the bounded nature of probability estimates.

For a given experimental dataset, we generate bootstrap samples by parametric resampling. Specifically, for estimated probabilities $\hat{p}$ obtained from $N$ measurements, we draw $M$ bootstrap replicates from the binomial distribution:
\begin{equation}
    n_k^* \sim \mathrm{Binomial}(N, \hat{p}), \quad k = 1, \ldots, M.
\end{equation}
Each replicate yields a bootstrap estimate $\hat{p}_k^* = n_k^* / N$. The full data processing pipeline, including gate sequence analysis, fidelity and SPAM estimation, is applied identically to each bootstrap sample.

The $(1-\alpha)$ confidence interval is then constructed from the empirical quantiles of the bootstrap distribution:
\begin{equation}
    \mathrm{CI}_{1-\alpha} = \left[ \hat{\theta}^*_{(\alpha/2)}, \, \hat{\theta}^*_{(1-\alpha/2)} \right],
\end{equation}
where $\hat{\theta}^*_{(q)}$ denotes the $q$-th quantile of the bootstrap distribution $\{\hat{\theta}_k^*\}_{k=1}^M$ for a parameter of interest $\theta$ (gate fidelity or SPAM error probability).

For all results presented in this work, we use $M = 1000$ bootstrap samples and report $95\%$ confidence intervals ($\alpha = 0.05$). The asymmetric error bars are computed as
\begin{equation}
    \sigma^+ = \hat{\theta}^*_{(0.975)} - \tilde{\theta}, \qquad \sigma^- = \tilde{\theta} - \hat{\theta}^*_{(0.025)},
\end{equation}
where $\tilde{\theta}$ is the mean value of the bootstrap distribution.

This approach guarantees that confidence intervals lie within $[0, 1]$.

\end{document}